# Frequent Patterns mining in time-sensitive Data Stream

Manel ZARROUK[1], Mohamed Salah GOUIDER[2]

[1] University of Gabès. Higher Institute of Management of Gabès
6000 Gabès, Gabès, Tunisia
*zarrouk.manel@gmail.com*

[2] University of Tunis. Higher Institute of Management of Tunis
2000 Le Bardo, Tunis, Tunisia
*ms.gouider@yahoo.fr*

**Abstract**

Mining frequent itemsets through static Databases has been extensively studied and used and is always considered a highly challenging task. For this reason it is interesting to extend it to data streams field. In the streaming case, the frequent patterns' mining has much more information to track and much greater complexity to manage.

Infrequent items can become frequent later on and hence cannot be ignored. The output structure needs to be dynamically incremented to reflect the evolution of itemset frequencies over time.

In this paper, we study this problem and specifically the methodology of mining time-sensitive data streams. We tried to improve an existing algorithm by increasing the temporal accuracy and discarding the out-of-date data by adding a new concept called the "Shaking Point". We presented as well some experiments illustrating the time and space required.

## 1. Introduction

The Data is a small word leading to an enormous computing-field and vital-ubiquitous component of technology's life. Since this notion is so important, we have to exploit it in a very accurate way by analyzing the huge amount or "mines" of collected data through time or differently said Data Mining. Recently, new generation of data has appeared and it's called data streams.

A Data Stream is an infinite and continuous sequences of data received at a high-speed rate and which disable the capability of storing them in memory for a later processing and this because of their historical dimension of real time. This concept has been defined in many references [11, 6, 23 and 17].

Since traditional DBMS can't fulfill the data stream requirement, they have been revised to give birth to the new Data Stream Management System. (fig.1)

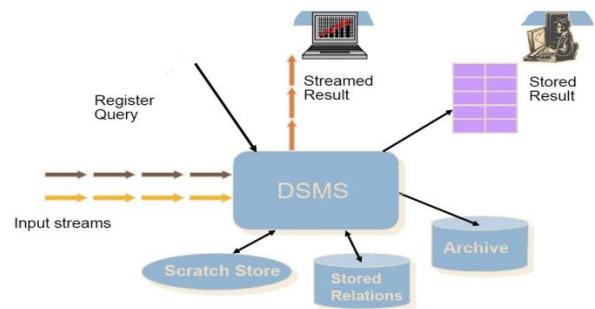

Fig1. A DSMS model

A number of methods have been used within these management systems and have proved being powerful in solving many problems in extracting knowledge from streaming information as: Data stream clustering [1, 2, 15], Data stream classification [1, 10, 18], frequent patterns mining [4, 14, 19], Change detection in data streams [3, 9, 20], Stream cube analysis of multi-dimensional streams [7] and Dimensionality reduction and forecasting [25,27].

Yet, between these tasks, the most challenging one in the research field is the frequent patterns mining which focuses on discovering frequently occurring patterns from different type of datasets (unstructured, semi-structured and structured).

In this paper, we study this problem and specifically the methodology of mining time-sensitive data streams which is previously studied by J. Han et al. in their paper published on November 2003 titled "Mining Frequent Item-sets over Arbitrary Time Intervals in Data Streams" [12]. We tried to improve the temporal accuracy and to discard the out-of-date data by adding a new concept called the "Shaking Point".

The remaining of the paper is structured as follows: section 2 introduces more deeply the concept of the frequent patterns mining and its related works. The modifications applied to an existing algorithm are detailed in section 3. In section 4 some experiments are presented. Finally, section 5 highlights some points which can be discussed and used as future research tracks and concludes the paper.

## 2. Frequent Patterns Mining in Data Stream

Mining or discovering frequent pattern itemsets consist on the first step of association rule mining process. This stage is accomplished by extracting all frequent itemsets that meet a support threshold given by the user. The second step is around creating rule associations that fulfill a confidence criterion based on the frequent itemsets found earlier. Since this stage can be applied in a simple way, a big part of the researches were done around the first one "how to discover efficiently all frequent itemsets in a data stream".

Some of the algorithms proposed under this method are: Lossy counting algorithm BTS [22], DSM-FI algorithm [21], EstDEC[26], FP-Streaming [12], Instant [24], Est-win [28], Moment [8], etc.

Within this paper, we focused on the time-sensitive aspect of the frequent patterns mining, as long as in many cases changes of patterns and their trends are more interesting than patterns themselves when the goal is to find changes or evolution of frequent patterns through time [12].

The algorithm FP-Streaming introduced in [12] satisfies this specificity. In their algorithm, frequent patterns are actively maintained under a tilted-time window framework in order to answer time-sensitive queries. The frequent patterns are compressed and stored using a tree structure similar to FP-tree [16] and updated incrementally with incoming transactions. The FP-tree provides a base structure to facilitate mining in a static batch environment. In the concerned algorithm, a transaction-tree is used for storing transactions for the current time window. On the same time, a similar tree structure, called pattern-tree, is used to store frequent patterns in the past windows. Our time-sensitive stream mining model, FP-stream, includes two major components: (1) pattern-tree, (2) tilted-time window [16].

For these reasons we chose this algorithm as a foundation of our work and tried to improve it in purpose to increase its itemsets temporal accuracy and discard the obsolete data from the FP-Stream output structure to fasten the response of the queries. The improvements are explained in the next section.

## 3. Modifications applied in the FP-Streaming model

### 3.1. Discarding the Tail Pruning and increasing accuracy

As introduced in [12], the tail pruning lightens the size of the FP Stream structure and decreases the space requirement of the algorithm. But as a result of this type of pruning, we no longer have an exact temporal frequency of a specified item $x$ over, rather an approximate frequency $\hat{f}_x(T)$ [13].

So due to this drawback of the Tail Pruning and its impact to the temporal frequencies of itemsets, we decided in this work to discard it so we will have a good precision in the temporal frequency of each item and its exact behavior graph over time.

This type of pruning is applied on the FP Stream, which is the output structure of our algorithm. This output tree structure is stored on the main memory to be consulted to answer the user queries. By discarding the Tail Pruning, the FP Stream structure will need much more space. We allowed this change in the algorithm despite the increasing space requirement, because it doesn't affect the algorithm performance (since it is not about the input structure which is the transaction tree) so we won't face any over fitting case. And since the memory is almost getting costless by years[5], the memory required for storing the FP Streaming will not pose a problem and will serve the goal of increasing the temporal accuracy of the user queries answers.

### 3.2. Stream generation and input stream management

In this algorithm, the input stream is generated from a data stream generator. This generator creates a stream of random items with random length separated by an (x) mark in a fast and constant rate. This stream replaces the datasets used to experiment algorithms.

The algorithm reads the input stream and fills in the batch within a period θ fixed by the user. So the window length will be dependant of θ .At the end of this time period, the algorithm ignores the stream and proceeds with analyzing the batch. Other while, between each two batches, the time elapsed is calculated and stored in a table to be used in need to know the time of stream lost without processing.

### 3.3. Leaf fading and elimination of the obsolete data

#### 3.3.1. The use of the Time-stamping in the original algorithm

To facilitate the insertion of zeros into tilted-time window tables, each table in the FP-stream has an associated time-stamp. When the table is created the time-stamp is set to the current time unit. When an insertion is made, the time stamp is reset to the current time unit. The use of time-stamps will become clear is the next paragraph.

### 3.3.2. Leaf fading concept

In the FP–Stream structure (the output structure), for every arriving batch, all the nodes are updated (incremental part of the algorithm). New nodes are added, frequencies of others updated.

If the itemset of a node occurs on the current batch, his current frequency is added on the tilted-time window of this node. If not, the tilted-time window will be updated by a zero.

After several batches arrive, some nodes that are updated consequently by zeros will be found. These nodes represent obsolete itemsets, since those itemsets were not frequent or sub-frequent since a while.
To reduce obsolete data, we will proceed as well:

- Associate to each node a variable called "fading-factor".
- Determine a user-set parameter called "fading-support".
- Modify the use of the time-stamping by eliminate the update of the time-stamp of a node whose tilted-time window is updated by a zero.
- After N batches arrival ( N is a user-defined parameter ) we execute a shaking point.
- Shaking point: calculate the fading factor for each node which represents the difference between its time-stamp and the current-time stamp (subtract the time-stamp for each node from the current time-stamp). If the fading-factor of a node is larger or equal to the fading-support, so drop the node and its entire supper-tree.
- In the shaking point we proceed from the root of the tree to the leaves (in depth), if a node of an item-set *I* is dropped then none of its supersets need to be examined. So the procedure of the shaking-point can prune its search and visit of the supersets of *I*. This property is obvious to avoid useless operations (if a parent node is pruned all the branch below will be pruned , since the frequency of itemsets in the leaves will not be updated if the frequency of the parent itemset were not ).

The choice of the fading-support is very important because it affects the spatial frequency precision of the answers, since according to this support the algorithm will drop faded leaves of outdated itemsets (obsolete data) in one hand, but can affect the spatial precision of the algorithm in the other hand.

### 3.3.3. Illustrative example

The first part of the algorithm returns us a stream of the frequent patterns with the same structure of the input stream.
With this stream we will update the FP Stream.

The structure of this stream is as following:
*(batch1)x((batch2)x(batch3)x ... (batch$_i$)*
The structure of a batch is as following:
(*itemset1)x(itemset2)x(itemset3) ... (itemset$_i$)*

In this example we will take a part of the stream which consists of the results of 3 batches:
- Frequent patterns of batch 1: *ACDxEVxACJxBFA*
- Frequent patterns of batch 2: *EVDxAxBFC*
- Frequent patterns of batch 3: *EVDxBFAxAH*
i) With the frequent patterns of the first batch (ACD, EV, ACJ, BFA) we create our FP Stream structure as illustrated below (fig. 2).

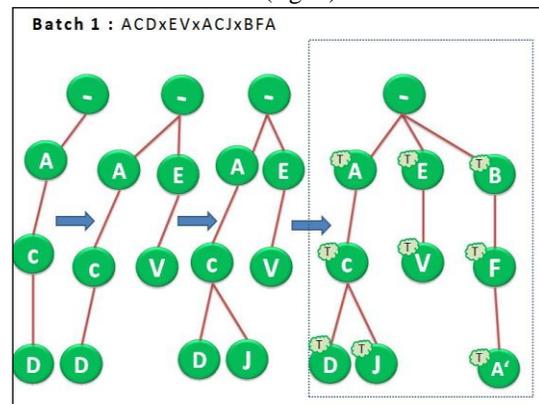

Fig 2. The FP Stream after the 1$^{st}$ batch update

⇨ All nodes are new-added
⇨ All time-stamps = T (current-system-time)

ii) We increment our FP Stream with the frequent patterns of the batch 2 which are EVD, A and BFC (fig. 3).

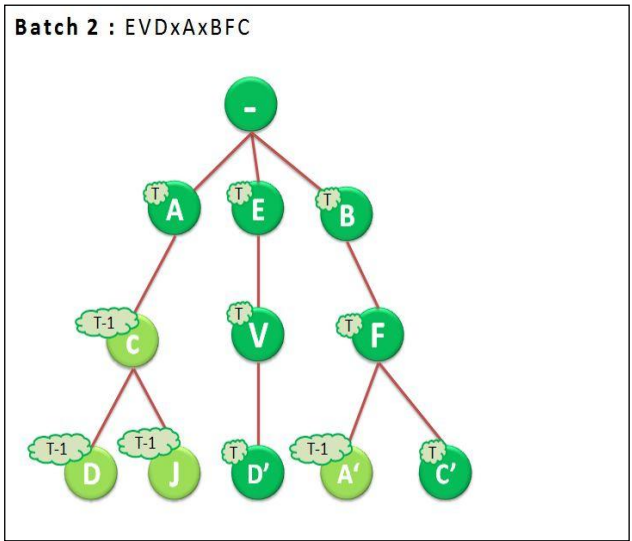

Fig 3. The FP Stream after the 2nd update

- ⇨ Item-sets which tilted-time window are updated with frequency ≠ 0: E,EV,A,B,BF
- ⇨ New itemsets : EVD',BFC'
- ⇨ Item-sets which tilted-time window are updated with frequency = 0: AC,ACD,ACJ,BFA'
- ⇨ Nodes with time-stamp updated to T : A,E,V,D',B,F,C'
- ⇨ Nodes faded with time-stamp not updated =T-1 : C,D,J,A'

iii) We increment the FP Stream as well by the frequent patterns of the third batch which are EVDC, BFA and AH. (fig. 4).

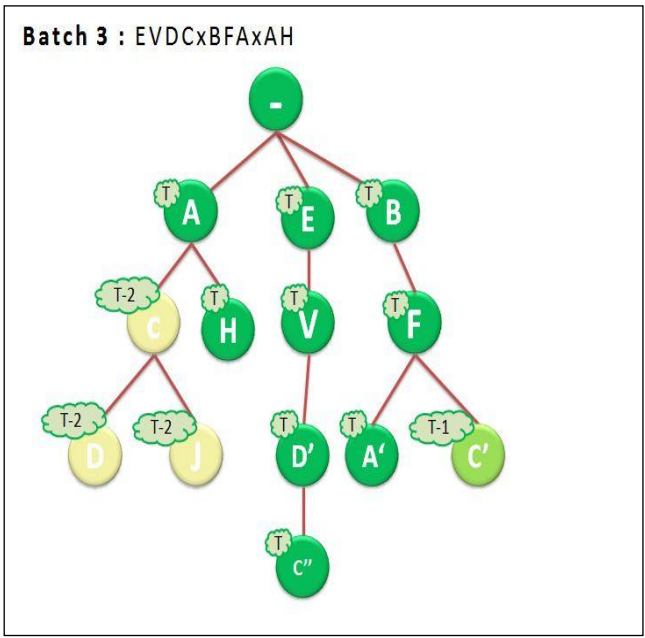

Fig 4. The FP Stream after the 3rd batch update

- ⇨ Item-sets which tilted-time window are updated with frequency ≠ 0: E, EV, EVD', B, BF, BFA', A.
- ⇨ New itemsets: AH, EVD'C''.
- ⇨ Item-sets which tilted-time windows are updated with frequency = 0: AC, ACD, ACJ, BFC'.
- ⇨ Nodes with time-stamp updated to T: A, E, V, D', B, F, A', H.
- ⇨ Nodes faded with time-stamp not updated: C, C', D, J.

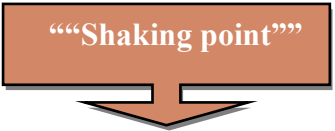
""Shaking point""

iv) In this stage, after N batches (N=3), we execute the "Shaking point" on our updated FP Stream with a fading-support=2. For a fading-factor ≥ 2 we drop the node (Fig5)

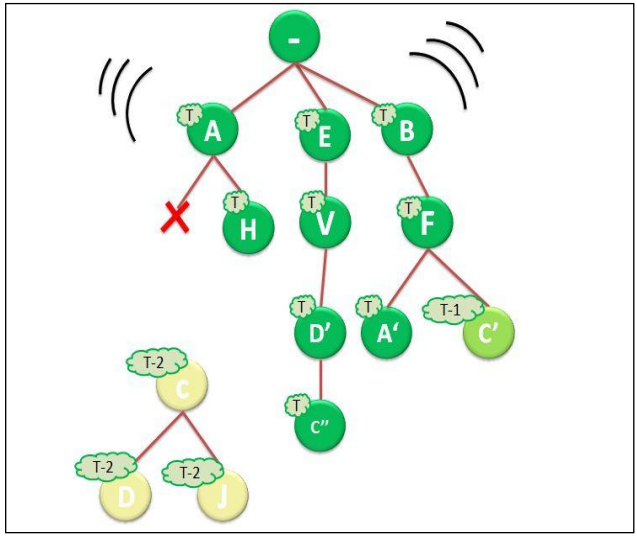

Fig 5. The Shaking Point

Proceed in depth-left:
$fading\_factor_{(A)} = T - T = 0$
$fading\_factor_{(C)} = T - (T - 2) = 2$ ➔ Drop (C) and all its subsets [(D) and (J)]
$fading\_factor_{(H)} = T - T = 0$
$fading\_factor_{(E)} = T - T = 0$
$fading\_factor_{(V)} = T - T = 0$
$fading\_factor_{(D')} = T - T = 0$
$fading\_factor_{(C'')} = T - T = 0$
$fading\_factor_{(B)} = T - T = 0$
$fading\_factor_{(F)} = T - T = 0$
$fading\_factor_{(A')} = T - T = 0$
$fading\_factor_{(C')} = T - (T - 1) = 1$

As a result, the FP-Stream will drop 3 obsolete-nodes and will be as shown below (fig. 6).

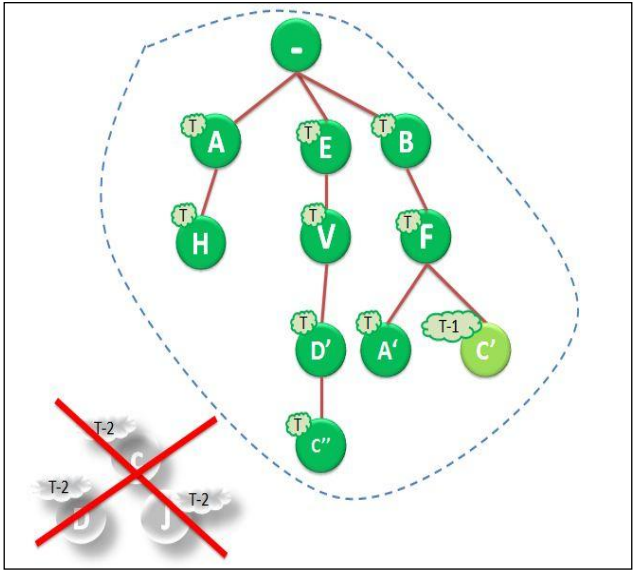

Fig 6. The FP Stream after the Shaking Point

### 3.4. Algorithm

The algorithm is presented in three parts. The first part (Algorithm1) ensures the management of the input stream under batches to provide to the next algorithm a stable batches environment to proceed. The second algorithm (Algorithm2) warrants the incremental update of the FP Stream structure with incoming batches. As for the third algorithm (Algorithm3), it maintains this structure by pruning it (shaking point) so the obsolete data which occupy an extra space without utility will be dropped.

**Input**: FP Stream structure, min-support $\sigma$, max-support-error $\epsilon$, incoming batch $B_i$ of transactions, lattice Offc, sliding window with a fixed period $W_t$, offline-time-counter, fading-factor in each node, fading-support $F_{supp}$
**Output**: FP Stream structure updated

*PS: the lines in **bold** present the amelioration
The lines between ## are discarded*

*Algorithm1: managing the incoming data stream*
*Method:*
- **Fill the batch $B_i$ during the window period which is $W_t$**
- **After $W_t$:**
  - **Stop receiving transactions**
  - **Launch the offline-time-counter to count the time of unprocessed transaction stream**
  - **Start proceeding the batch (*Algorithm 2*)**
- **If ( the batch $B_i$ is reset to empty) :**
  - **Stop offline-time-counter and stock its value in the lattice Offc**
  - **Proceed with *algorithm 2* in a loop**

*Algorithm 2: (FP-streaming) (Incremental update of the FP-stream structure with incoming stream data)*
*Method:*
- Initialize the FP-tree to empty.
- Sort each incoming transaction $t$, according to f-list, and then insert it into the FP-tree without pruning any items.
- When all the transactions in $B_i$ are accumulated, update the FP-stream as follows:
  - Mine itemsets out of the FP-tree using FP-growth algorithm in [16] modified as below. For each mined itemset $I$, check if $I$ is in the FP-stream structure.
    If $I$ is in the structure, do the following:
    - ✓ Add $f_I(B)$ to the tilted-time window table for $I$
    - ✓ *# # Conduct tail pruning ##*
    - ✓ *## If the table is empty, then FP-growth stops mining supersets of I (Type II Pruning). Note that the removal of I from the FP-stream structure is deferred until the scanning of the structure (next step). ##*
    - ✓ *## If the table is not empty, then FP-growth continues mining supersets of I ##*
    - ✓ Update the time-stamp of the node corresponding to the itemset $I$ to the current-system-time
    If $I$ is not in the structure do the following:
    - ✓ If $f_I(B) \geq \epsilon |B|$, insert $I$ into the structure (its tilted-time window will have only one entry $f_I(B_i)$) **and its time-stamp will be up-to-date)**
    - ✓ Else FP-Growth stops mining the supersets of $I$ (Type 1 pruning)
  - Scan the FP-stream structure (depth-first search). For each itemset $I$ encountered, check if $I$ was updated when $B$ was mined. If not, then insert 0 into $I$'s tilted-time window table **without updating the time-stamp** (did not occur in).
    *## Prune I's table by tail pruning. Once the search reaches a leaf, if the leaf has an empty tilted-time window table, then drop the leaf. If there are any siblings of the leaf,*

*continue the search with them. If there were no siblings, then return to the parent and continue the search with its siblings. Note that if all of the children of the parent were dropped, then the parent becomes a leaf node and might be dropped.##*

**Algorithm 3: Shaking point (dropping the out-of-date data)**
*Method:*
- **After N batches we proceed with the shaking-point as following:**
  - **Scan the FP-stream structure (depth-first search). For each itemset $I$ and with "CST" to refer to "current-system-time" do the following :**
  - **Calculate the Fading-Factor $Fad_{factor} = CST - I's\ timestamp$**
    - ✓ **If ( $Fad_{factor} \geq Fad_{supp}$ ) drop the $I$'s node and all its supersets**
    - ✓ **Else proceed and check the supersets**

## 4. Performance study and experiments

In this section, we report our performance study. We describe first our experimental set-up and then our results.

### 4.1. Experimental setups

Our algorithm was written in Java and compiled using Eclipse java indigo. All our experiments were performed on a PC equipped with a 3.22 GHz Intel Core i5 and a 4 Go main memory. The operating system was Windows 7 premium familial edition. All experiments were run without any other user on the machine.

The stream data was generated by a synthetic data generator coded in the algorithm. This generator creates a stream of random items with random length separated by an (x) mark in a fast and constant rate.

### 4.2. Experimental results

We performed 4 sets of experiments. $\sigma$ was fixed respectively at 0.2 , 0.4 , 0.6 , 0.8 (per cent). In all sets of experiments the data stream was fed into the program from the synthetic-stream-generator. The size of sliding window and consecutively the batch is fixed to the number of transactions arriving for 5 seconds.

At each batch the following statistics were collected: the total number of milliseconds required per batch "runtime" (which does not include the time of reading transaction from the input stream) and the size of the FP-Stream structure at the end of each batch in bytes (does not include the temporary FP-tree structure used for mining the batch). In all graphs presented, the $x$ axis represents the batch number and min-support refers to $\sigma$.

Figures 7 and 8 present the time and size requirements results respectively.

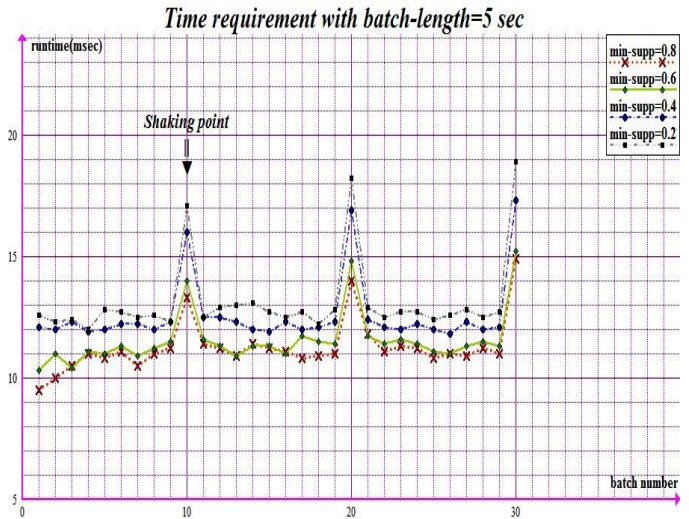

Fig 7. Algorithm's time requirement

In fact, it's obvious that the minimum support and the required time to process have a proportional relation, when the minimum support increases the run-time decreases. But we notice as well that in the graph there is a spike (peak) where the run-time of the program increases in a remarkable way. This spike is due to the Shaking-point that occurs after each 10 batches in our example, and this since this shaking-point is checking the entire FP-Stream structure, it's straightforward that it will use a more extra time to proceed.

So what we can conduct without fail from this graph that the modified algorithm has the same behavior toward the variation of the minimum support as the original FP Streaming algorithm [14].

The most important requirement which is that the algorithm does not fall behind the stream (the FP-stream structure must be updated before the next batch of transactions is processed) is always handled in our case because the batch won't receive new transactions unless the FP-tree is set to empty after updating the FP Stream.

This specification will lead to a loss of some of the streaming data which is inevitable in the data stream mining field. Moreover considerable improvements can be met by reducing the runtime of our algorithm which leads to the reduction of the loss of streaming.

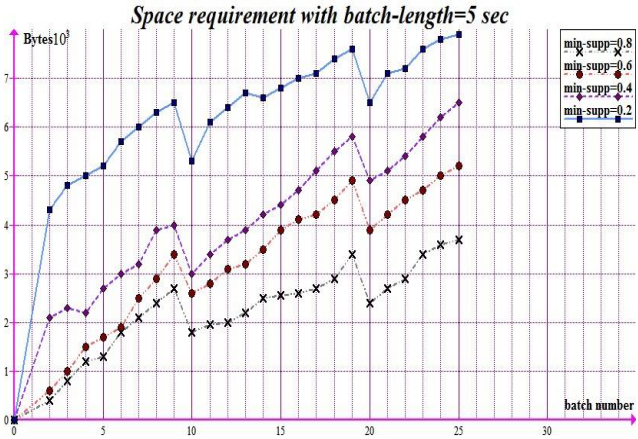

Fig 8. Algorithm's space requirement

Even in the space requirement graph we notice that the higher the minimum support is, the less size the FP Stream occupies. This aspect is normal since when increasing the min-supp we apply a bigger constraint on the items to meet the minimum support required so the much less frequent itemsets will be found.

We can mark a narrow through (drop) in the space usage at the 10th and 20th batch in some recursive way. This behavior breaks down the forward growing up of the graph. Again, this strike is explained by the occurring of the Shaking point every 10 batches for this example. The Shaking point, by dropping some obsolete nodes, is reducing the size of the FP Stream structure and is slowing down its fast growth through time.

## 5. Conclusions

In this paper, we introduced the huge field of mining data streams, its environment and its first submergence. We took a brief look to its methods and algorithms as well to focus finally on an approach to mine time-sensitive frequent patterns on different time granularities. This model is based on an effective pattern-tree structure called FP-Stream, which consists of an in-memory frequent/sub-frequent pattern tree with tilted-time window embedded. Efficient algorithms are devised for constructing, maintaining, and updating an FP-stream structure over data streams. Moreover, we focused on the updating of the incremental part of the algorithm and tried to contribute in a way that increases the temporal frequency of the results and eliminates the outdated data. Efficiency was evaluated by several experimentations of the proposed method which demonstrate that time sensitive frequent data can be maintained through a stream environment dependless on available main memory.

Some aspects still can be discussed in this algorithm and can lead to another future works:

- **Spatial accuracy approximation:** By applying the Shaking points, the global frequency of an itemset can be slightly affected which leads to an approximate spatial frequency. But in the other side we keep the temporal frequency accurate by discarding the Tail Pruning.

- **Query answering:** we focused this work on the study and the amelioration of the incremental part of the algorithm (FP Stream structure). However, there's an important side of the subject which worth a future study, this part is the query answering from the FP Stream tree.

- **Loss reduction:** the loss of the transaction while proceeding with the algorithm is saved in the OffC Lattice. This loss must be discounted or by finding a more intelligent behavior of the sliding window or by reducing the run time of the algorithm.